\documentclass[aps,prb,twocolumn,superscriptaddress,showpacs,reprint]{revtex4-1}
\usepackage{graphicx}
\usepackage{mathrsfs}
\usepackage{bm}
\usepackage{amsmath}
\usepackage{multirow}
\usepackage{bigstrut}
\DeclareMathOperator{\sgn}{sgn}

\begin{document}

\title{Topological phases in gated bilayer graphene: Effects of Rashba spin-orbit coupling and exchange field}
\author{Zhenhua Qiao}
\affiliation{Department of Physics, The University of Texas at Austin, Austin, Texas 78712, USA}
\author{Xiao Li}
\affiliation{Department of Physics, The University of Texas at Austin, Austin, Texas 78712, USA}
\author{Wang-Kong Tse}
\affiliation{Department of Physics, The University of Texas at Austin, Austin, Texas 78712, USA}
\author{Hua Jiang} \affiliation{International Center for Quantum Materials, Peking University, Beijing 100871, China}
\author{Yugui Yao} \affiliation{School of Physics, Beijing Institute of Technology, Beijing 100081, China}
\author{Qian Niu} \affiliation{Department of Physics, The University of Texas at Austin, Austin, Texas 78712,
USA}\affiliation{International Center for Quantum Materials, Peking University, Beijing 100871, China}

\begin{abstract}
We present a systematic study on the influence of Rashba spin-orbit
coupling,  interlayer potential difference and exchange field on the
topological properties of bilayer graphene. In the
presence of only Rashba spin-orbit coupling and interlayer potential
difference, the band gap opening due to broken out-of-plane inversion
symmetry offers new possibilities of realizing tunable topological phase
transitions by varying an external gate voltage. We find a two-dimensional $Z_2$
topological insulator phase and a quantum valley Hall phase in $AB$-stacked bilayer graphene and
obtain their effective low-energy Hamiltonians near the Dirac
points. For $AA$ stacking, we do not find any topological insulator
phase in the presence of large Rashba spin-orbit coupling. When the exchange field
is also turned on, the bilayer system exhibits a rich variety of
topological phases including a quantum anomalous Hall phase, and we obtain the phase diagram as a function of the Rashba spin-orbit
coupling, interlayer potential difference, and exchange field.
\end{abstract}

\date{\today}
\maketitle
\section{Introduction}

Topological insulator~\cite{TI-review} (TI) is a new phase of quantum
matter in materials with strong spin-orbit coupling. The quantum spin Hall
effects in graphene~\cite{Kane} and HgTe quantum wells~\cite{HgTe}
represent the first examples of two-dimensional topological
insulators. So far, semiconductor heterostructures like HgTe~\cite{HgTe-exp} and
InAs/GaSb~\cite{RuiruiDu} quantum wells offer the only realistic materials with strong spin-orbit coupling
that can realize the quantum spin Hall phase, and the intrinsic spin-orbit coupling in graphene was shown to be too
weak. ~\cite{WeakSOC}   Due to graphene's
attractiveness as a potential electronic material for emerging
nanotechnology, artificially enhancing the spin-orbit coupling
strength in graphene can open up new possibilities in graphene-based
spintronics, and several theoretical and experimental works have
addressed the effects of enhanced spin-orbit coupling in graphene by doping with heavy adatoms such as indium or
thallium ~\cite{RuqianWu}, doping with 3d/5d transition metal
atoms~\cite{HongbinZhang,qiao1,ding}, and interfacing with metal substrates, e.g., Ni(111)~\cite{RashbaSOC}.

Due to band gap opening from broken out-of-plane inversion symmetry, gated bilayer graphene is a quantum valley-Hall insulator (QVHI) characterized by a quantized valley Chern number. In our recent work,~\cite{qiao2} we have reported that the presence of Rashba
spin-orbit coupling turns the gated bilayer graphene system from a
QVHI into a $Z_2$ TI, with the phase boundary
given by $\lambda^2_R=U^2+t^2_{\perp}$ where $\lambda_R$, $U$ and
$t_{\perp}$ denote the strengths of Rashba spin-orbit coupling,
interlayer potential difference and interlayer tunneling amplitude
respectively. In this paper, we obtain low-energy effective
Hamiltonians for the topological insulator phase, valid for small $U \ll \lambda_R$ and below the
topological phase transition, as well as for the quantum valley Hall
phase above the phase transition. In the presence of different Rashba
spin-orbit coupling strengths on the top and bottom layers ($\lambda_R^1 \neq\lambda_R^2$), we show that
the topological insulator phase remains robust as long as
$\lambda_R^{1} \lambda_R^{2} > t^2_\perp$. When the time-reversal
symmetry is broken by an additional exchange field $M$, the bilayer
system hosts different topological phases characterized by different
Chern numbers $\mathcal{C}=2,4\,{\rm sgn}(M)$ and valley Chern numbers
$\mathcal{C}_v=2,4\,{\rm sgn}(U)$, and the phase boundaries associated
with the topological phase transitions are given by $U=\pm M$ and
$U^2+t^2_\perp-M^2-\lambda^2_R=0$.

The rest of this paper is organized as
follows. Section~{\ref{section1}} introduces the tight-binding and low-energy effective Hamiltonian of
$AB$-stacked bilayer graphene in the presence of Rashba spin-orbit
coupling, exchange field, and interlayer potential difference. In
Section~{\ref{section2}} we obtain the low-energy effective
Hamiltonians of the $Z_2$ TI and the quantum
valley Hall phases. We consider in Section~{\ref{section3}} the case of $AA$-stacked bilayer
graphene. The effect of different Rashba spin-orbit coupling strengths
on the top and bottom layers of $AB$-stacked bilayer graphene is
considered in Section~{\ref{section4}}. Finally, in
Section~{\ref{section5}} we obtain the phase diagram of $AB$-stacked
bilayer graphene as a function of $\lambda_R$, $U$ and $M$.

\section{System Hamiltonian}\label{section1}
The tight-binding Hamiltonian of $AB$-stacked bilayer
graphene in the presence of Rashba spin-orbit coupling, exchange
field, and interlayer potential difference is
given by ~\cite{James,qiao2}
\begin{eqnarray}
H_{\rm BLG}&=&H^T_{\rm SLG}+H^B_{\rm SLG}+t_\bot \sum_{i\in T, j \in
B}{ c^\dagger_{i}c_{j}} \nonumber \\
&&+{{U}}\sum_{i\in T}{ c^\dagger_{i}c_{i}}-{{U}}\sum_{i\in B}{ c^\dagger_{i}c_{i
}}\label{BilayerH},
\end{eqnarray}
where the first two terms $H^{T,B}_{\rm SLG}$ denote the monolayer
graphene Hamiltonian for the top (T) and bottom (B) layer (see below),
the third term represents the tunneling Hamiltonian that couples the
top and bottom layers, and the last two terms take into account a
potential difference $2U$ between the top and bottom layers. The
single-layer graphene Hamiltonian is:~\cite{qiao1,Kane,DonnaSheng,qiao3}
\begin{eqnarray}
H_{\rm SLG}=H_0+H_{{R}}+H_{{M}},
\end{eqnarray}
with
\begin{eqnarray}
&&H_0=- t \sum_{\langle{ij}\rangle}{ c^\dagger_{i}c_{j}}; \nonumber \\
&&H_{{R}}={i} t_{{R}}\sum_{\langle{ij}\rangle; \alpha, \beta
}\hat{\mathbf{e}}_{z}{\cdot}({{s}_{\alpha
\beta}}{\times} {\mathbf{d}}_{ij})c^\dagger_{i \alpha } c_{j \beta };   \nonumber \\
&&H_{{M}}={{M}} \sum_{i;\alpha,\beta}{ c^\dagger_{i\alpha}{s}^z_{\alpha\beta}c_{i\beta}}, \nonumber
\end{eqnarray}
where $\langle...\rangle$ runs over all the nearest neighbor
sites with hopping amplitude $t=2.6$\,eV, ${\bm s}$ are spin Pauli matrices with $\alpha$ and $\beta$
denoting up spin or down spin, and $c^\dag_{i\alpha}$ ($c_{i\alpha}$) is the
electron creation (annihilation) operator on site $i$. $H_R$ describes
the Rashba spin-orbit coupling with coupling strength $t_R$ and
${\mathbf{d}}_{ij}$ is a lattice vector pointing from sites $j$ to
$i$, and $H_M$ the exchange field contribution with magnetization $M$.

In the momentum space ~\cite{note1}, we perform an expansion of the
momentum about the valley points $K$ and $K'$ and obtain the following eight-band effective Hamiltonian:~\cite{James,qiao2}
\begin{align}
H=& v (\eta \sigma_x k_x + \sigma_y k_y )\bm{1}_s\bm{1}_\tau +
\frac{t_\bot}{2}  ( \sigma_x \tau_x- \sigma_y \tau_y) \bm{1}_s
\nonumber \\
+&\frac{\lambda_{{R}}}{2}(\eta\sigma_x s_y - \sigma_y s_x)
\bm{1}_\tau
+ {{M}} s_z \bm{1}_\sigma \bm{1}_\tau +{U} \tau_z \bm{1}_s \bm{1}_\sigma,
\label{EightBandH}
\end{align}
where $\eta =\pm1$ label the valley degrees of
freedom, $\bm{\sigma}$ and $\bm{\tau}$ are Pauli matrices
representing the $AB$ sublattice and top-bottom layer degrees of freedom, $\bm{1}$ is a
$2\times2$ identity matrix. The bare graphene Fermi velocity is given
by $v = 3at/2$ with $a$ the lattice constant and Rashba spin-orbit
coupling is given by $\lambda_{{R}} = 3t_{{R}}$. For simplicity, we set the lattice constant $a$ to be unity henceforth.

\section{Four-band effective Hamiltonian}
\label{section2}
\begin{figure}[!]
\includegraphics[width=7cm,angle=0]{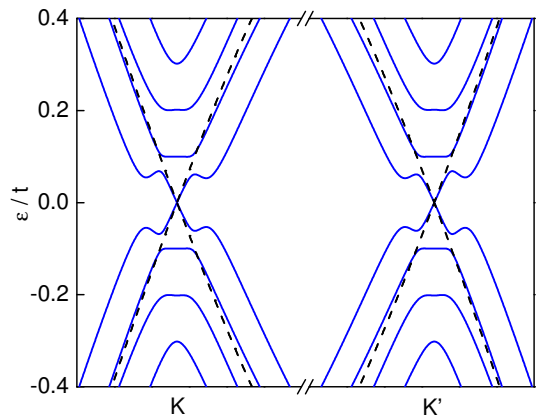}
\caption{(Color online) Solid line: bulk band structure of bilayer
  graphene at the phase transition point between the quantum
  valley-Hall and the $Z_2$ topological insulator phases. The parameters
  used are $U=0.10t$, $t_\perp=0.143t$ and $t_R=0.058t$. Dashed line:
  bulk band structure of pristine single layer graphene. The band gap
  closing at phase transition occurs precisely at the valleys $K$ and $K'$.} \label{DiracComaprison}
\end{figure}
In Ref.~[\onlinecite{qiao2}], we reported numerical tight-binding  calculations
showing that $AB$-stacked bilayer graphene under external interlayer
potential undergoes a topological phase transition from a QVHI to a two-dimensional $Z_2$ TI.
Figure~\ref{DiracComaprison} shows that, at the phase transition
point, the bulk band gap of the bilayer graphene system obtained from
Eq.~(\ref{EightBandH}) closes at the valley points at the zero energy (for comparison Figure~\ref{DiracComaprison} also shows the
band structure of a pristine single-layer graphene). It is therefore
possible to obtain low-energy Hamiltonian descriptions near band gap
closing, which occurs when bilayer graphene turns into a TI from a semi-metal for small $U$ and at the phase transition
point $U = U_0$ between TI and QVHI. In the following, we expand the eight-band Hamiltonian
Eq.~(\ref{EightBandH}) up to the leading order in momentum $\bm{k}$ around the valley points and obtain
a reduced four-band effective Hamiltonian that captures the low-energy
physics of the system near phase transitions.
\begin{figure}[!]
\includegraphics[width=6cm,angle=0]{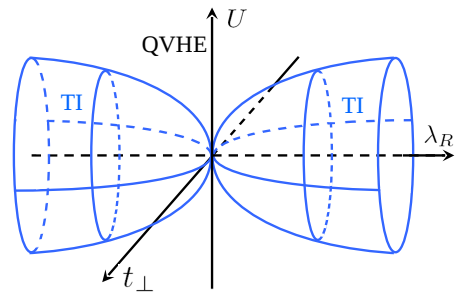}
\caption{(Color online) Phase diagram of bilayer graphene as a
  function of ${\mathrm{U}}$, $\lambda_R$, $t_\perp$. The region
  inside the ellipsoid corresponds to the $Z_2$ topological insulator
  state and outside region to the quantum valley Hall insulator phase. The phase boundary is given by Eq.~\eqref{Eq:PhaseBoundary}} \label{wholephasediagram}
\end{figure}

\begin{figure}[!]
\includegraphics[width=8cm,angle=0]{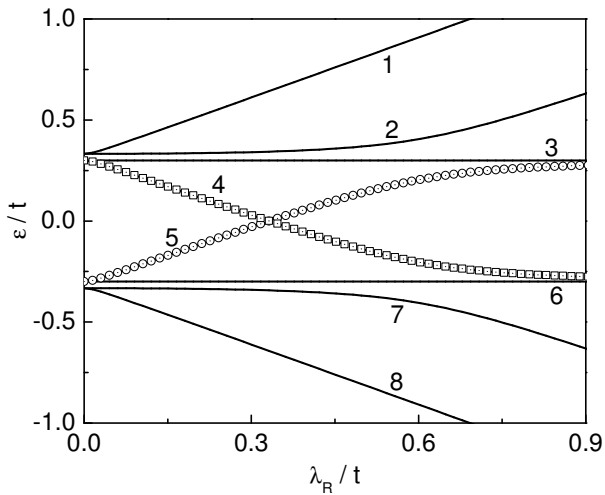}
\caption{Energy dispersion at valley $K$ (${\bf k}=0$) as a function of
  Rashba spin-orbit coupling $\lambda_{\mathrm{R}}$ at fixed
  ${\mathrm{U}}=0.3t$ and $t_\perp=0.143t$. One notices that band inversion occurs between band `4' and band `5' at the critical
  Rashba spin-orbit coupling value $\lambda_R \simeq0.33t$.}  \label{Invertband}
\end{figure}

We assume equal Rashba spin-orbit coupling strengths in both top
and bottom layers in this Section  and address the effects of unequal
Rashba strengths in Section~{\ref{section4}}. Figure~\ref{Invertband}
illustrates our results for the energy bands obtained numerically from
the eight-band Hamiltonian in Eq.~(\ref{EightBandH}).
We observe that the fourth band and the fifth band become inverted
when the Rashba strength increases beyond a critical value signaling a
topological phase transition. Imposing ${\bm {k}}=0$ in
Eq.~\eqref{EightBandH} gives the condition
for gap closing of the bulk bands as~\cite{qiao2}
\begin{eqnarray}
\lambda_{{R}}^2={{U}}^2+t_{\bot}^2.\label{Eq:PhaseBoundary}
\end{eqnarray}
The (${\mathrm{U}}$, $\lambda_R$, $t_\perp$) phase space is therefore
divided into two regions as illustrated in
Fig.~\ref{wholephasediagram}: the system is in the $\mathcal{C}_v=4 \sgn(U)$ QVHI
phase when $\lambda_{{R}}^2 < {{U}}^2+t_{\bot}^2$, while it is in the $Z_2$
TI phase characterized by $\mathcal{C}_v=2\sgn(U)$
and $Z_2=1$ when $\lambda_{{R}}^2 > {{U}}^2+t_{\bot}^2$.

\begin{figure}[!]
\includegraphics[width=8cm,angle=0]{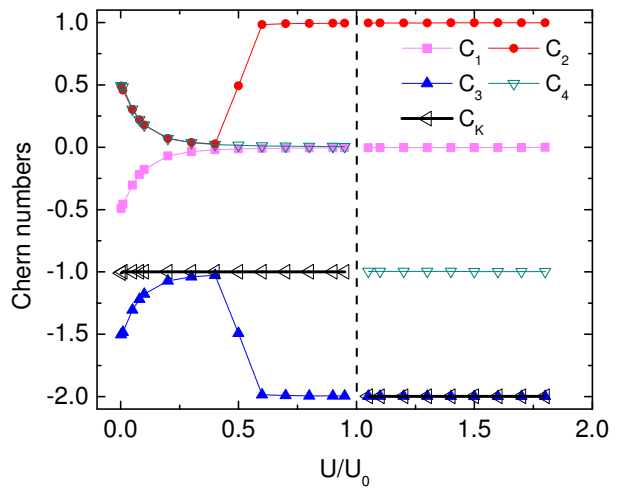}
\caption{(Color online) Chern number contribution from each valence
  band obtained from Eq.~\ref{EightBandH} at valley $K$ as a
  function of $U/U_0$. The Rashba spin-orbit coupling and interlayer
  tunneling are fixed. $U_0$ is the critical value (see the dashed
  line) of interlayer potential separating the 2D TI and QVH phases. For $U<U_0$, the total Chern number $\mathcal{C}_K=-1$, but the contribution from each valence band varies as a function of $U/U_0$. For $U>U_0$, the total Chern number $\mathcal{C}_K=-2$ and the contribution from each valence band is independent of $U/U_0$.} \label{ChernNumberDependence}
\end{figure}

In Fig.~\ref{ChernNumberDependence}, we plot the Chern number
contributions from each valence band using the eight-band Hamiltonian
in Eq.~(\ref{EightBandH})
near the valley $K$. Here, we fix the Rashba spin-orbit coupling and
interlayer coupling as $\lambda_R/t=0.617$ and $t_\perp/t=0.143$. The
topological phase transition point occurs at $U_0 \equiv
\sqrt{\lambda^2_R-t^2_\perp}$. In the TI phase when
$U<U_0$, the plot shows that the contribution to the total Chern number from
each valence band varies as a function of $U$, and in particular there
are two regimes where the Chern number contributions are distributed differently
among the bands.  For $U\rightarrow0$,
$C^{K}_{1,2,3,4}=-0.5,0.5,-1.5,0.5$ while for $U\rightarrow U^{-}_0$,
$C^{K}_{1,2,3,4}=0.0,1.0,-2.0,0.0$. For the QVHI phase
occurring when $U>U_0$, the Chern number contribution from each
valence band is constant as a function of $U$ with $C^{K}_{1,2,3,4}=0.0, 1.0, -2.0, -1.0$.
In the following we study the low-energy physics of the TI and QVHI states in these three regimes.

\subsection{Near Semimetal-TI Phase Boundary: $U\rightarrow 0$ and $U\ll \lambda_{R}$}
In the basis \{$A_{1\uparrow}$, $B_{1\downarrow}$, $A_{2\uparrow}$,
$B_{2\downarrow}$, $A_{1\downarrow}$, $B_{1\uparrow}$,
$A_{2\downarrow}$, $B_{2\uparrow}$\}, the eight-band Hamiltonian Eq.~(\ref{EightBandH}) can be written at valley $K$ as:
\begin{eqnarray}
H_{\mathrm{K}}=\left[
\begin{array}{ccccccccc}
H_1 & T \\
T & H_2
\end{array}
\right]
\end{eqnarray}
with
\begin{eqnarray}
H_1=\left[
\begin{array}{ccccccccc}
+U & 0 & 0 & 0\\
0 & +U & 0 & 0 \\
0 & 0 & -U & 0 \\
0 & 0 & 0 & -U
\end{array}
\right],
\end{eqnarray}
\begin{eqnarray}
H_2=\left[
\begin{array}{ccccccccc}
+U & i \lambda_{\mathrm{R}} & 0 & 0\\
-i \lambda_{\mathrm{R}} & +U & 0 & 0 \\
0 & 0 & -U & i \lambda_{\mathrm{R}} \\
0 & 0 & -i \lambda_{\mathrm{R}} & -U
\end{array}
\right],
\end{eqnarray}
and
\begin{eqnarray}
T=\left[
\begin{array}{ccccccccc}
0 & v k_- & 0 & t_\perp \\
v k_+ & 0 & 0 & 0 \\
0 & 0 & 0 & v k_- \\
t_\perp & 0 & v k_+ & 0
\end{array}
\right].
\end{eqnarray}
In the limit $U \rightarrow 0$ and $U\ll \lambda_{R}$, $H_1$ and $H_2$
correspond respectively to the lower bands [i.e. $\varepsilon=\pm U$] and higher bands [i.e. $\varepsilon= \pm (\lambda_{\mathrm{R}} \pm U)$]. In the vicinity of $K$, the coupling $T$ between $H_1$ and $H_2$ becomes very weak. Therefore, the original eight-band Hamiltonian can be reduced to an effective four-band Hamiltonian:\cite{note0}
\begin{eqnarray}
H^{{\mathrm{eff}}}_{\mathrm{K}}&\simeq&H_1-TH^{-1}_2 T \nonumber \\
&=&\frac{1}{\lambda_R}\left[
\begin{array}{cccc}
U{\lambda_R} & {i v^2 k^2_-} &  0 & {2i  t_\perp v k_-} \\
{-i v^2 k^2_+} & U{\lambda_R} & 0 & 0 \\
0 & 0 & -U{\lambda_R} & {i v^2 k^2_-} \\
{-2i  t_\perp v k_+} & 0 & {-i v^2 k^2_+} & -U{\lambda_R}
\end{array}
\right]. \nonumber \\ \label{effctiveFormK1}
\end{eqnarray}

Similarly, the eight-band Hamiltonian at valley $K^\prime$ in the basis of \{$A_{1\downarrow}$, $B_{1\uparrow}$, $A_{2\downarrow}$, $B_{2\uparrow}$, $A_{1\uparrow}$, $B_{1\downarrow}$, $A_{2\uparrow}$, $B_{2\downarrow}$\} can be expressed as:
\begin{eqnarray}
H_{\mathrm{K^\prime}}=\left[
\begin{array}{ccccccccc}
H_1 & T_0 \\
T_0 & H_2
\end{array}
\right]
\end{eqnarray}
with
\begin{eqnarray}
T_0=\left[
\begin{array}{ccccccccc}
0 & -v k_+ & 0 & t_\perp \\
-v k_- & 0 & 0 & 0 \\
0 & 0 & 0 & -v k_+ \\
t_\perp & 0 & -v k_- & 0
\end{array}
\right].
\end{eqnarray}

Using the formula of Eq.~\eqref{effctiveFormK1}, the resulting reduced four-band Hamiltonian can be written as:
\begin{eqnarray}
H^{{\mathrm{eff}}}_{\mathrm{K^\prime}}\simeq \frac{1}{\lambda_R} \left[
\begin{array}{cccc}
U{\lambda_R} & {i v^2 k^2_+} &  0 & {-2i  t_\perp v k_+} \\
{-iv^2 k^2_-} & U{\lambda_R} & 0 & 0 \\
0 & 0 & -U{\lambda_R} & {i v^2 k^2_+} \\
{2i  t_\perp v k_-} & 0 & {-i v^2 k^2_-} & -U{\lambda_R}
\end{array}
\right].\label{effctiveFormK2}
\end{eqnarray}

Upon diagonalization of Hamiltonians in Eqs.~(\ref{effctiveFormK1}) and (\ref{effctiveFormK2}), the energy dispersions at both $K$ and $K'$ can be obtained and share the same form:
\begin{eqnarray}
\varepsilon= \pm\frac{\sqrt{\lambda^2_RU^2+\epsilon^4_k+2\epsilon^2_kt^2_\perp \pm 2\epsilon^2_k\sqrt{\lambda^2_R U^2 +t^4_\perp + \epsilon^2_k t^2_\perp}}}{\lambda_R}, \nonumber \\ \label{reducedHamiltonian}
\end{eqnarray}
where $\epsilon_k=vk$.

\begin{figure}[!]
\includegraphics[width=7cm,angle=0]{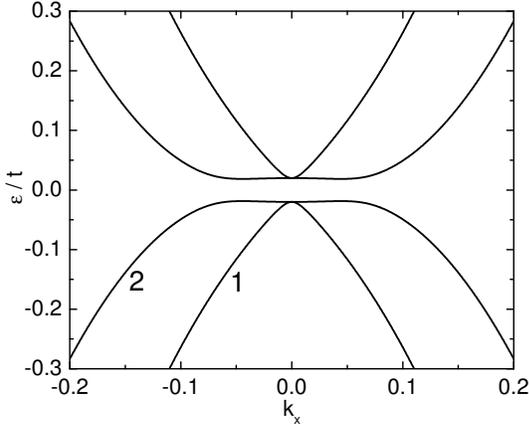}
\caption{Bulk band structure of the effective four-band model in the limit of $\lambda_R/U\gg1$ along the profile of $k_y=0$. `1' and `2' label the two valence bands. Here, we choose $\lambda_R/=0.20t$ and $U=0.02t$. Note that the two conduction or valence bands touch at $k=0$.} \label{EnergyBands-Eff}
\end{figure}

Figure~\ref{EnergyBands-Eff} plots the bulk band structure of the
four-band effective Hamiltonian in Eq.~(\ref{reducedHamiltonian})
along $k_y=0$. One can see that a bulk band gap opens
and the resulting conduction and valence bands touch at
$k_x=0$. The bulk band gap opening signals an insulating state. To
reveal its topological property, we have evaluated the Berry phase
contribution from the occupied valence bands below the band gap. In a
 continuum model Hamiltonian, the Chern number is calculated by
 integrating the Berry curvature in the entire momentum space:
\begin{eqnarray}
\mathcal{C}=\frac{1}{2\pi}\sum_{n=1,2}\int^{+\infty}_{-\infty}\int^{+\infty}_{-\infty}
d k_x d k_y \Omega _n(k_x,k_y), \label{ChernNumber}
\end{eqnarray}
where $\Omega _n(k_x,k_y)$ is the momentum-space Berry curvature at
($k_x$, $k_y$) of the $n$-th band, and is given by
\begin{eqnarray}
\Omega_n(\bm{k})=-{\sum_{n^{\prime} \neq n}} {\frac{2 {\rm {Im}}
\langle \psi_{n \bm{k}}|v_x|\psi_{n^\prime \bm{k}} \rangle \langle
\psi_{n^\prime \bm{k}}|v_y|\psi_{n \bm{k}} \rangle }
{(\omega_{n^\prime}-\omega_{n})^2}}, \label{berry}
\end{eqnarray}
where $\omega_n\equiv \varepsilon_n/ \hbar$, and $v_{x(y)}$ is the
velocity operator along the $x$($y$)-direction.

\begin{figure}[!]
\includegraphics[width=8.5cm,angle=0]{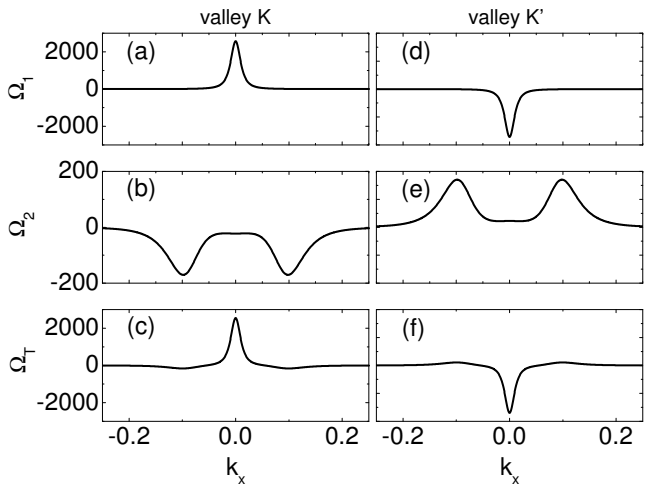}
\caption{Berry curvature distribution $\Omega$ at valleys $K$ and $K'$
  along $k_y=0$ of the effective four-band model in the limit of $\lambda_R/U\gg1$. Upper panels: Berry curvature distribution $\Omega_1$ for the lowest valence band labeled as `1' in Fig.~\ref{EnergyBands-Eff}. Middle panels: Berry curvature distribution $\Omega_2$ for the valence band close to the band gap labeled as `2' in Fig.~\ref{EnergyBands-Eff}. Lower panels: the total valence band Berry curvature distribution $\Omega_T$. The parameters adopted here are the same as those used in Fig.~\ref{EnergyBands-Eff}.} \label{BerryCurvature-Eff}
\end{figure}

In Fig.~\ref{BerryCurvature-Eff}, we display the Berry curvature
distribution $\Omega$ along $k_y=0$ for both valleys $K$ and $K'$. One
observes that the Berry curvatures are exactly opposite at the two
valleys $K$ and $K'$ for each band. In particular, we find that the
total Berry curvatures around $K$ or $K'$ do not share the same sign
in the whole momentum space, in contrast with the Berry curvature
distribution in the quantum anomalous Hall effect in single layer graphene~\cite{qiao1} or the conventional QVHI in graphene due to the presence of staggered AB sublattice potential.\cite{ding,DiXiao}

By numerically evaluating the integration in the momentum space, the
Chern numbers at valleys K and K' are found to be
\begin{eqnarray}
\mathcal{C}_K=-\mathcal{C}_{K'}={\sgn} (U), \label{ChernNumberResult}
\end{eqnarray}
in which the Chern number contribution from each valence band are
\begin{eqnarray}
\mathcal{C}^1_K&=&-\mathcal{C}^1_{K'}=-\frac{1}{2} {\sgn} (U), \\
\mathcal{C}^2_K&=&-\mathcal{C}^2_{K'}=+\frac{3}{2} {\sgn} (U),
\end{eqnarray}
where the supscripts label the valence band indices in Fig.~\ref{EnergyBands-Eff}.

From the principle of bulk-edge correspondence, one expect that there
is only one pair of edge states propagating along the system
boundaries. Since time-reversal symmetry is preserved in the our
system, one concludes that this nontrivial insulating state belongs
to the $Z_2$ TI class. Therefore, the edge modes
are robust against weak non-magnetic impurities. Moreover, since
different valleys are encoded into the counter-propagating edge
channels,~\cite{qiao2} they are further protected by the large
momentum separation as long as inter-valley scattering is
forbidden. As a consequence, these edge modes are also robust against
smooth non-magnetic and magnetic impurities. From Eq.~(\ref{ChernNumberResult}), the valley Chern number of this $Z_2$ TI state is:
\begin{eqnarray}
\mathcal{C}_v=
{\mathcal{C}_K-\mathcal{C}_{K'}}=\frac{2e^2}{h} {\sgn} (U),
\end{eqnarray}
which is half of that in the conventional QVHI.\cite{James}

\subsection{Near TI-QVHI Phase Boundary: $U\rightarrow \sqrt{\lambda^2_R - t^2_\perp}$}
In the following discussion, we set $\lambda_R$ and $t_\perp$ as fixed,
and allow $U$ to slightly deviate from the phase transition point
$U_0$, i.e., $U=U_0+\Delta$, where $U_0=\sqrt{\lambda^2_R -
  t^2_\perp}$, $\Delta > 0 \,\mathrm{and}\,< 0$ correspond
respectively to the QVHI and TI phases.

When the topological phase transition occurs, the bulk band gap closes
at the valley points $K/K'$. Hence, at the critical value $U_0$
and $\bm{k} = 0$, the system Hamiltonian in the basis of $\{B_{1\downarrow}, A_{2\uparrow}, A_{1\uparrow}, A_{2\downarrow}, B_{2\uparrow}, B_{2\downarrow}, B_{1\uparrow}, A_{1\downarrow} \}$ can be expressed as
\begin{eqnarray}
H=\left[
\begin{array}{cccccccc}
U_0 &  0   & 0       & 0           & 0          & 0       & 0           & 0 \\
0   & -U_0 & 0       & 0           & 0          & 0       & 0           & 0 \\
0   &  0   & U_0     & 0           & t_\perp    & 0       & 0           & 0 \\
0   &  0   & 0       & -U_0        & -i\lambda_R& 0       & 0           & 0 \\
0   &  0   & t_\perp & i \lambda_R & -U_0       & 0       & 0           & 0 \\
0   &  0   & 0       & 0           & 0          & -U_0    & 0           & t_\perp\\
0   &  0   & 0       & 0           & 0          & 0       &      U_0    & i\lambda_R\\
0   &  0   & 0       & 0           & 0          & t_\perp & -i\lambda_R & U_0
\end{array} \label{Ham_U0}
\right].
\end{eqnarray}
The eigenenergies can be obtained from the above as
$\varepsilon=0,0,\pm U_0, \pm \varepsilon_1, \pm \varepsilon_2$, where
$\varepsilon_1=- U_0/2 + \sqrt{8\lambda^2_R + U^2_0}/2$ and
$\varepsilon_2=- U_0/2 - \sqrt{8\lambda^2_R + U^2_0}/2$. The former
four correspond to the low-energy part, while the latter four
correspond to the high-energy part. Based on our analysis of the Berry
curvature from the tight-binding model, the valley Chern number arises
only from the low-energy bands and the high-energy bands give no contribution.

The unitary transformation matrix that diagonalizes Eq.~\eqref{Ham_U0}
is presented in the Appendix. After some manipulations, the effective Hamiltonian can be
obtained from
\begin{eqnarray}
H_{\rm eff}=H_P-T H^{-1}_Q T^{\dagger},
\end{eqnarray}
where explicit expressions of $H_P, T, H_Q$ are also given in the
Appendix. The effective Hamiltonian to first order in $\bm k$ is then given by
\begin{eqnarray}
H^{(1)}_{\rm eff}=
\left[
\begin{array}{cccccccc}
-\frac{U^2_0}{\lambda^2_R}\Delta & -i\frac{t_\perp}{\lambda_R} k_- & 0 & -\frac{U_0}{\sqrt{2}\lambda_R} k_+ \\
i\frac{t_\perp}{\lambda_R} k_+ & \frac{U^2_0}{\lambda^2_R} \Delta & \frac{U_0}{\sqrt{2}\lambda_R}k_- &0 \\
0& \frac{U_0}{\sqrt{2}\lambda_R}k_+ &\Delta + U_0 &0 \\
-\frac{U_0}{\sqrt{2}\lambda_R}k_- &0&0&-\Delta - U_0 \\
\end{array} \nonumber
\right], \label{eff-FirstOrder}
\end{eqnarray}

\begin{figure}[!]
\includegraphics[width=8.5cm,angle=0]{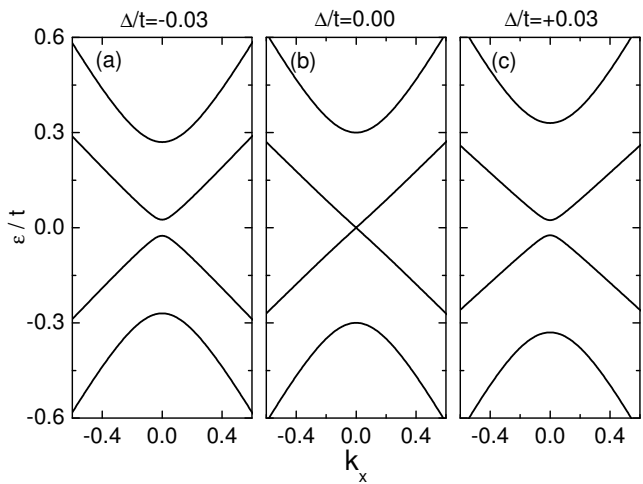}
\caption{Band structure obtained from the effective Hamiltonian
  Eq.~({\ref{eff-FirstOrder}}) along $k_y=0$. The parameters for the
  phase transition point are set to be $U_0/t=0.30$ and
  $\lambda/t=0.33$. (a) $\Delta/t=-0.03$; (b) $\Delta/t=0.00$; (c)
  $\Delta/t=+0.03$. The bulk band gap decreases toward zero when
  $\Delta/t=0.00$ and then reopens when $\Delta$ becomes positive.
 \label{Bands-eff2} }
\end{figure}

\begin{figure}[!]
\includegraphics[scale=0.45]{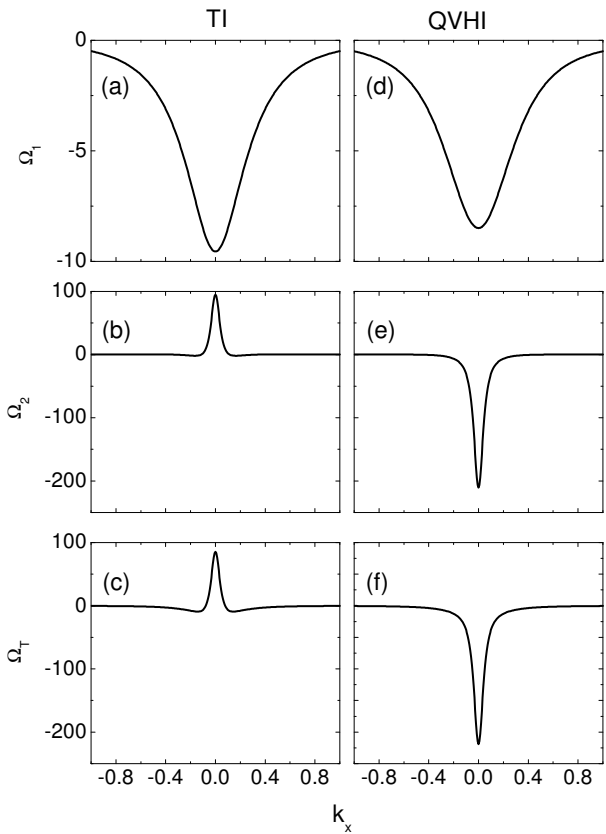}
\caption{Berry curvature distribution $\Omega$ along $k_y=0$ for $\Delta=\pm 0.03$. Other parameters are $U_0/t=0.30$ and $\lambda/t=0.33$. (a)-(c): Berry curvature distribution for each valence bands  and the summary of all valence bands at $\Delta/t=-0.03$. (d)-(f) Berry curvature distribution for each valence bands  and the summary of all valence bands at $\Delta/t=+0.03$.
\label{BerryCurvature-eff2}}
\end{figure}

In the following, we show that the first-order effective Hamiltonian
at valley $K$ is sufficient to capture the topological phase
transition between the QVHI and TI phases. Figure~\ref{Bands-eff2}
displays the band structures along $k_y=0$ for three different values
of $\Delta/t=-0.03,0.00,0.03$ at fixed $U_0/t=0.30$ and
$\lambda_R/t=0.33$. One observes that when $\Delta$ increases from
negative to positive, closing and reopening of the bulk band gap occur
as expected. Figure~\ref{BerryCurvature-eff2} plots the Berry curvatures along
$k_y=0$ of the two valence bands below the band gap and the total
Berry curvature for $\Delta/t=\pm0.03$. The Berry curvatures for the
first valence band $\Omega_1$ in both cases are similar sharing the
same sign [panels (a) and (d)]. For the second valence band, the Berry
curvatures for $\Delta/t=-0.03$ exhibit both positive and negative
signs [panel (b)], but for $\Delta=+0.03$, they are both negative
[panel (e)]. As a consequence, the total berry curvatures at
$\Delta/t=-0.03$ include both positive and negative contributions
[panel (c)], and those at $\Delta=+0.03$ share the same negative
sign. By using Eq.~(\ref{ChernNumber}), the Chern numbers acquired by
the valence bands at valley $K$ is evaluated as
$\mathcal{C}_{K}=-1,-2$ for $\Delta=-0.03,+0.03$
respectively. Following a similar derivation, one obtains the
effective Hamiltonian at valley $K'$ with the corresponding Chern
numbers $\mathcal{C}_{K'}=1,2$ for $\Delta=-0.03,+0.03$
respectively. In this way, the valley Chern number in the topological
insulator phase is
$\mathcal{C}_{v}=\mathcal{C}_{K}-\mathcal{C}_{K'}=-2$ while that in
the quantum valley Hall phase is $\mathcal{C}_{v}=\mathcal{C}_{K}-\mathcal{C}_{K'}=-4$.

In all the three limits, the valley-Chern numbers from the resulting four-band effective Hamiltonians are consistent with those from the direct eight-band full Hamiltonian.

\section{AA-stacked bilayer graphene}\label{section3}
Bilayer graphene is composed of two monolayers of graphene, usually
arranged in $AB$ or $AA$ stacking pattern. In previous Sections, we
have predicted a $Z_2$ TI phase in the $AB$ stacking
configuration. It therefore becomes a natural question to ask whether the $AA$ stacking configuration
can also host a $Z_2$ TI phase. In this Section, we demonstrate that the $AA$-stacked bilayer graphene
does not realize a $Z_2$topological insulator state in the presence of Rashba spin-orbit coupling and interlayer potential.

\begin{figure}[!]
\includegraphics[width=8.5cm,angle=0]{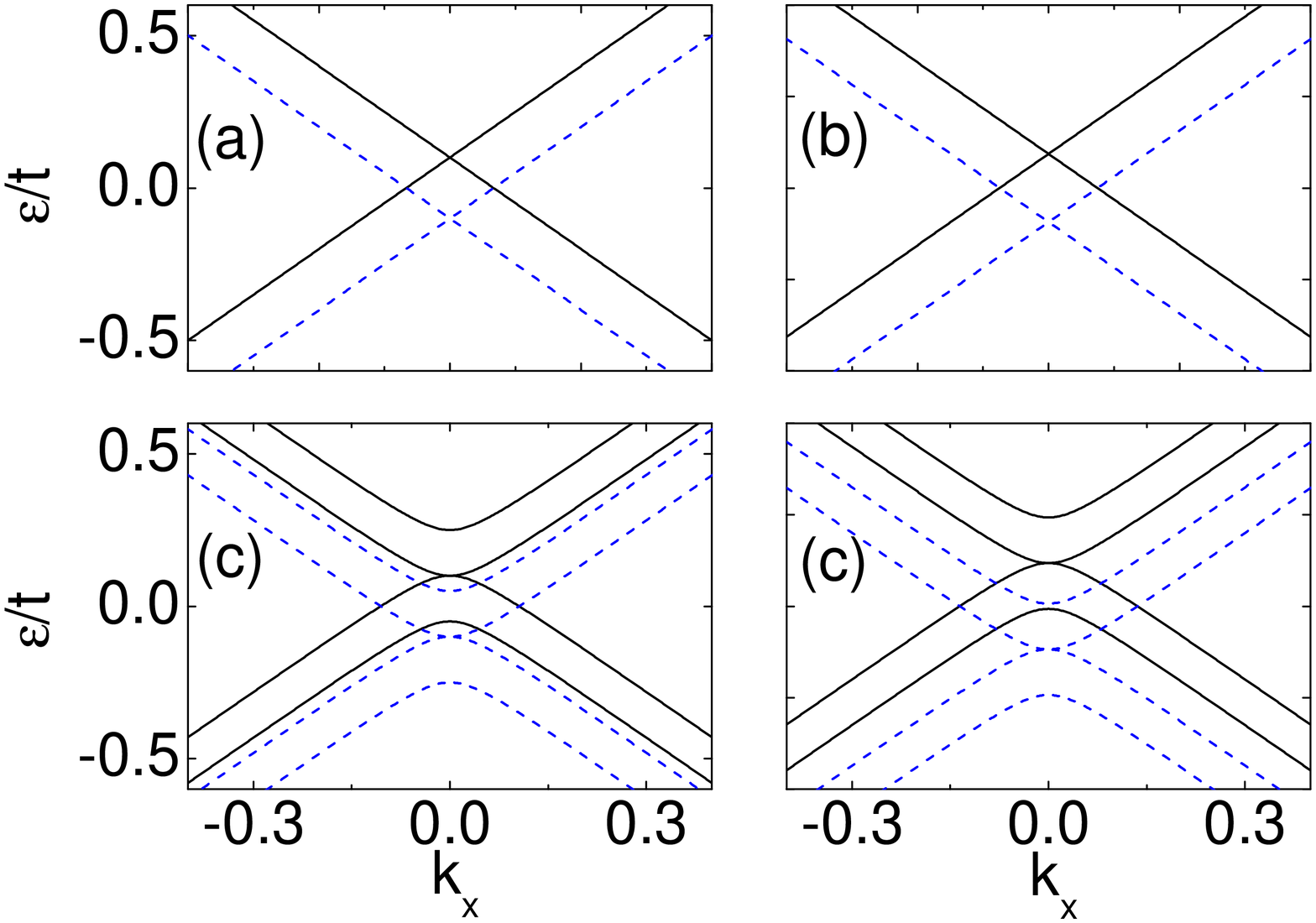}
\caption{(Color online) Bulk band structures of $AA$-stacked bilayer graphene along $k_y=0$ at valley $K$. (a) Pristine graphene, $U/t=0$ and $\lambda_R/t=0$; (b) $U=0.10t$ and $\lambda_R=0$; (c) $U=0$ and $\lambda_R=0.15t$; (d) $U=0.10t$ and $\lambda_R=0.15t$. Solid and dashed bands correspond to the top and bottom layers, respectively. Here the interlayer coupling is set to be $t_\perp=0.10t$.} \label{bandsAA}
\end{figure}

Figure~\ref{bandsAA} plots the bulk band structures of $AA$-stacked
bilayer graphene along $k_y=0$ obtained from the low-energy continuum model Hamiltonian at valley $K$.~\cite{note2} For pristine AA-stacked graphene, the linear Dirac
dispersion near valley $K$ still holds as shown in panel (a),
resembling two copies of monolayer graphene with a relative shift of
$2t_\perp$ (solid and dashed lines are used to label the bands from
top and bottom layers). In the presence of an interlayer potential
difference, there is no bulk band gap opening [see
panel (b)] since the inversion symmetry with respect to the graphene
plane is not violated. If only the Rashba spin-orbit coupling is turned on, one finds that
again the resulting band structure is a combination of two copies of the
monolayer graphene's band structures with a relative shift [see panel
(c)]. When both Rashba spin-orbit coupling and interlayer potential
difference are present, no bulk band gap appears.
We therefore conclude that inversion symmetry breaking is a necessary
requirement for the $Z_2$ TI in the bilayer graphene system.
In addition to $AA$ and $BB$ stacking, twisted bilayer graphene presents another
possibility which has attracted much
recent interest.~\cite{twist1,twist2,twist3,twist4,twist5} In
future works, it will be interesting to study the possibility of
inducing a $Z_2$ TI in a twisted bilayer graphene.

\section{Effects of different Rashba spin-orbit couplings on two layers}\label{section4}
We have studied the $Z_2$ TI state while assuming
the same Rashba spin-orbit coupling in the top and bottom layers of
the $AB$-stacked bilayer graphene. In bilayer graphene, Rashba spin-orbit coupling is
extremely weak, and one has to employ external means to enhance
the Rashba spin-orbit coupling, e.g., doping it with heavy metal atoms or
interfacing it with substrates. Therefore, the resulting Rashba
spin-orbit couplings are likely to be different on the top and bottom
layers. In the following, we discuss the effect of different top and bottom Rashba
spin-orbit coupling strengths on the resulting TI state.

We adopt the low-energy continuum Hamiltonian
Eq.~\eqref{EightBandH} and assume different values of top and bottom
Rashba spin-orbit couplings $\lambda_1\neq\lambda_2$.
We first consider the case when both $\vert\lambda_{1}-\lambda_2\vert$
and $U$ are small. In this case, we found from our numerical calculations that although the
conduction and valance bands are no longer symmetric with respect to
$\varepsilon=0$, the bands still close exactly at the valley points
$K$ and $K'$; it is thus possible to obtain an analytic formula that
describes the band closing condition. After imposing
${\bm k}=0$, one finds that two of the eigenenergies are $\varepsilon=\pm U$, while the remaining six ones satisfy the following equations:
\begin{widetext}
\begin{equation}
\varepsilon^3+(-1)^{i}U\varepsilon^2-(\lambda_i^2+U^2+t_\perp^2)\varepsilon
-(-1)^{i}U(U^2-\lambda_i^2+t_\perp^2)=0, i=1,2. \label{Eq:unevenRashba}
\end{equation}
\end{widetext}

We search for the condition when the top of the valence band and the
bottom of the conduction band touch, closing the bulk band gap at the
$K$ or $K'$ point. This can be translated into the condition that the two equations in Eq.~\eqref{Eq:unevenRashba} have a common
real-valued solution that lies between $-U$ and $U$. The latter
condition is necessary in order to rule out the scenario that two
higher (lower) bands touch at the $K$ point.

\begin{figure}
\includegraphics[width=8cm,angle=0]{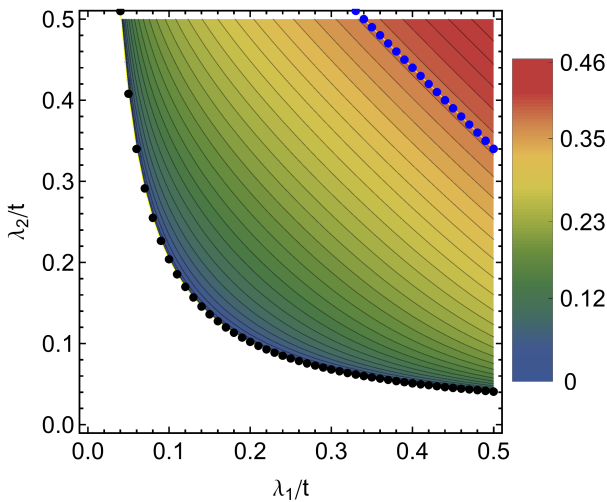}
\caption{(Color online) The interlayer potential $U$ that closes the band
  gap at the valley K point as a function of Rashba spin-orbit couplings
  $\lambda_1$ and $\lambda_2$, as predicted by
  Eq.~\eqref{Eq:NewPhaseBoundary}. Colors represent the amplitude of
  $U$. It can be seen that while such contours are hyperbolas for
  small $U$ (see the dotted line in the lower part of the graph), they
  become linear for larger $U$ (see the straight lines in the upper-right corner). In addition, the constraint of $\lambda_1\lambda_2>t_\perp^2$ is clearly demonstrated in this plot.} \label{Fig:NewPhaseBoundary}
\end{figure}

It turns out that the numerical search for a common solution is not as
simple as the case with identical Rashba effects in
Ref.~[\onlinecite{qiao2}], where we can directly require the common
solution to be $\varepsilon=0$. In the present case, however, the band
closing point is no longer fixed at $\varepsilon=0$, which makes it
difficult to obtain a simple analytical solution. Instead, we opt to solve
for $\lambda_i$ (and not for $\varepsilon$) from
Eq.~\eqref{Eq:unevenRashba}, obtaining
\begin{eqnarray}
\lambda_1^2&=&\dfrac{U-\varepsilon_0}{U+\varepsilon_0}(U^2+t_\perp^2-\varepsilon_0^2), \label{eqation1} \\
\lambda_2^2&=&\dfrac{U+\varepsilon_0}{U-\varepsilon_0}(U^2+t_\perp^2-\varepsilon_0^2), \label{eqation2}
\end{eqnarray}
where $\varepsilon_0$ is the common solution of the two equations in Eq.~\eqref{Eq:unevenRashba}. Then the band-closing point can be analytically obtained by dividing Eq.~\eqref{eqation1} by Eq.~\eqref{eqation2}:
\begin{eqnarray}
\varepsilon_0 = \dfrac{\lambda_2-\lambda_1}{\lambda_2+\lambda_1}U.
\end{eqnarray}

In order to have a band gap closing, these parameters must also satisfy the following condition
\begin{align}
t_\perp^2+\dfrac{4\lambda_1\lambda_2}{(\lambda_1+\lambda_2)^2}U^2=\lambda_1\lambda_2, \label{Eq:NewPhaseBoundary}
\end{align}
which is derived by multiplying Eqs.~\eqref{eqation1} and
\eqref{eqation2}. It is reassuring to see that when
$\lambda_1=\lambda_2$, this condition does reduce to the one given in
Eq.~\eqref{Eq:PhaseBoundary}. 
One can also rewrite Eq.~\eqref{Eq:NewPhaseBoundary} by expressing the interlayer potential difference $U$ as a function of $\lambda_1$ and $\lambda_2$:
\begin{align}
U=\pm\dfrac{\lambda_1+\lambda_2}{2}\sqrt{1-\dfrac{t_\perp^2}{\lambda_1\lambda_2}},
\end{align}
where we see that the band gap at the $K$ or $K'$ point will not be able to close if $\lambda_1\lambda_2<t_\perp^2$.

Figure~\ref{Fig:NewPhaseBoundary} plots the interlayer potential
difference $U$ in the ($\lambda_1$, $\lambda_2$) plane that satisfies
the band gap closing condition Eq.~\eqref{Eq:NewPhaseBoundary}. Colors
represent the strength of $U$.
In the blank region, no band gap
closing occurs  under the constraint of $\lambda_1\lambda_2>t_\perp^2$.
In the limit of small potential difference $U$, the contours of $U$
behave as hyperbolas given by $\lambda_1\lambda_2=t_\perp^2$ (see the
black dotted line), while in the large $U$ limit, the contours tend to straight lines given by $\lambda_1+\lambda_2=2U$.

\begin{figure}
\includegraphics[width=8.cm,angle=0]{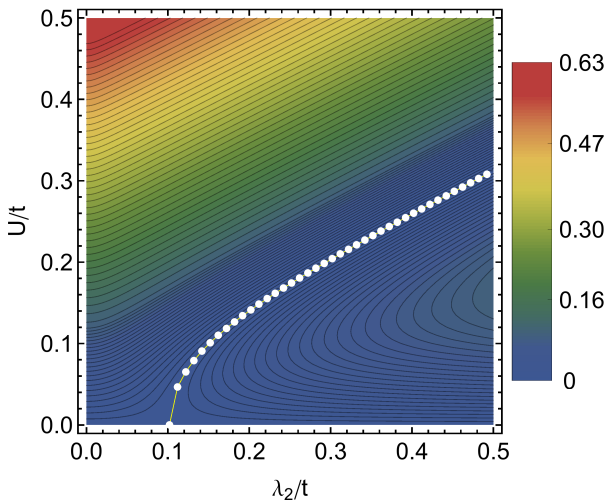}
\caption{(Color online) Numerically obtained band gap at valley $K$
  point as a function of the interlayer potential $U$ and bottom-layer
  Rashba
  spin-orbit coupling $\lambda_2$ at a fixed
  top-layer Rashba spin-orbit coupling $\lambda_1/t=0.2$. Color
  measures of the size of the band gap. The white dots are plotted
  from Eq.~\eqref{Eq:NewPhaseBoundary}. The analytical condition
  Eq.~\eqref{Eq:NewPhaseBoundary} for band gap closing overlaps with
  the numerical result. \label{Fig:PhaseDatSameLambda}}
\end{figure}

\begin{figure}
\includegraphics[width=8cm,angle=0]{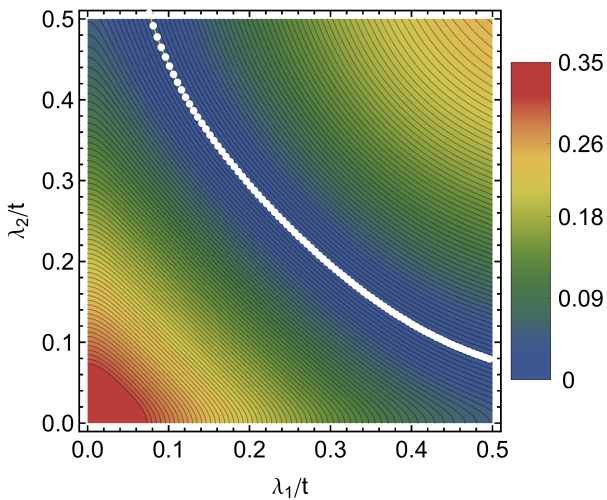}
\caption{(Color online) Numerically computed band gap at valley $K$
  point as a function of Rashba spin-orbit couplings in both top and
  bottom layers $\lambda_1$ and $\lambda_2$. In this plot the
  interlayer potential difference is set as $U=0.20t$. Color
  measures the size of the band gap in units of $t$. The white dots
  are plotted from Eq.~\eqref{Eq:NewPhaseBoundary}. The analytical
  result for the band gap closing condition agrees well with that
  obtained from numerical calculations.}\label{Fig:PhaseDatSameU}
\end{figure}

To verify the correctness of the analytical expression of the band gap
closing in Eq.~\eqref{Eq:NewPhaseBoundary}, we compare it with band
gap results
from direct numerical diagnolization of the eight-band continuum model Hamiltonian at valley $K/K'$. In Fig.~\ref{Fig:PhaseDatSameLambda}, we plot the band gap at $K$ point as a function of interlayer potential difference $U$ and Rashba spin-orbit coupling $\lambda_2$ at a fixed $\lambda_1=0.20t$. White dots
plot the analytic result Eq.~\eqref{Eq:NewPhaseBoundary}
corresponding to the boundary for band gap closing. We find that it
agrees well with the numerically obtained condition for band gap
closing. Similarly, in Fig.~\ref{Fig:PhaseDatSameU}, we
plot the band gap at $K$ point as a function of the two different
Rashba spin-orbit couplings $\lambda_1$ and $\lambda_2$ at a fixed
interlayer potential difference $U=0.2t$. The white dots obtained from
Eq.~\eqref{Eq:NewPhaseBoundary} fit exactly where the band gap closes
at the valley $K/K'$ point. Therefore, the analytical phase boundary in Eq.~\eqref{Eq:NewPhaseBoundary} indeed captures the band gap closing condition at valley $K/K'$.

For very different $\lambda_{1}$ and $\lambda_{2}$ and a large $U$,
we find that the conduction and valence bands can close
\textit{indirectly} at different momenta, and as a result there is no global
bulk gap even though the direct gap at the valley
points is nonzero. The global bulk gap is the smallest energy difference between the
conduction band and the valence band across the entire Brillouin zone.

\begin{figure}
\includegraphics[width=8cm,angle=0]{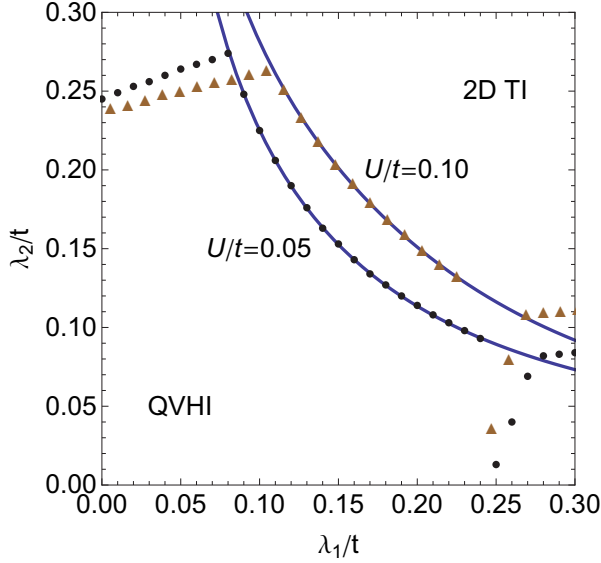}
\caption{(Color online) Comparison between the conditions for global bulk gap
  closing obtained numerically (circles and triangles) and local bulk gap closing predicted
  from Eq.~\eqref{Eq:NewPhaseBoundary} (solid line), for
  interlayer potential difference $U=0.05t$ and
  $U=0.10t$.}
\label{Fig:Globalgap}
\end{figure}

In Fig.~\ref{Fig:Globalgap}, we show the comparison between the
numerically computed band gap (circle or triangle) and direct band gap at the valley points
given by Eq.~\eqref{Eq:NewPhaseBoundary} (solid line).
One observes that for large differences in $\lambda_1$ and $\lambda_2$
and for large $U$, the numerically computed gap deviates from
Eq.~\eqref{Eq:NewPhaseBoundary}, indicating that band gaps close
indirectly at different momenta.

To examine the nontrivial topology of the phases before and after the
band gap closing, we calculate the valley-Chern numbers using the
continuum model and the $Z_2$ topological number using the
tight-binding model presented in Ref.~[\onlinecite{Z2}]. As depicted in
Fig.~\ref{Fig:Globalgap}, before the phase
transition, the system hosts a QVHI phase with $\mathcal{C}_v=\mathcal{C}_K-\mathcal{C}_{K'}=4$, while
after the phase transition, it enters into a TI phase with $Z_2=1$. Note that the TI state is
simultaneously a QVHI state characterized by
$\mathcal{C}_v=2$. These results are consistent with our findings in
the presence of identical Rashba spin-orbit couplings.~\cite{qiao2}
Therefore, we have shown that the $Z_2$ TI state we predicted in Ref.~[\onlinecite{qiao2}] remains robust when
the top and bottom layer Rashba spin-orbit coupling strengths become
different.

\section{Exchange field effect}\label{section5}
This section is dedicated to investigate the exchange field effect on
the QVHI state and $Z_2$ TI state in
gated bilayer graphene. In Ref.~[\onlinecite{qiao1}], we have found
that a bulk band gap will open in monolayer graphene in the presence of both Rashba spin-orbit coupling and exchange
field, inducing a quantum anomalous-Hall
phase~\cite{Haldane,ChaoxingLiu,Nagaosa,CongjunWu,FangZhong,GuangyuGuo,JiangHua}.
In Ref.~[\onlinecite{James}], we have shown that in gated bilayer
graphene, when the exchange field $M$ is larger than the interlayer
potential difference $U$, i.e. $M>U$, the system undergoes a
topological phase transition from the $\mathcal{C}_v=4$ QVHI phase into a $\mathcal{C}=4$ quantum anomalous-Hall
phase. In the following, we supplement this result with a new phase
boundary that has been overlooked in Ref.~[\onlinecite{James}].
We also discuss how the QVHI state with $\mathcal{C}_v=1,2$ evolves in the
presence of the exchange field.

For simplicity, we set the Rashba spin-orbit couplings to be the
same in both layers in this discussion. We start from the low energy
continuum model Eq.~(\eqref{EightBandH}), and consider the following
ingredients in our system: interlayer potential difference $U$, Rashba
SOC in both layers $\lambda_R$, and exchange field $M$. The conduction and valence bands are symmetric about $\varepsilon=0$,
and the bulk band gap closes at exactly the valley $K/K'$ point. The energy dispersion of the eight-band Hamiltonian at valley $K$  is determined by the following equations:
\begin{eqnarray}
&&\varepsilon = \pm(M - U),\nonumber \\
&&\lambda_R^2(M\pm\varepsilon+U)=(M\pm \varepsilon + U)[U^2+t_\perp^2-(M-\varepsilon)^2].\nonumber
\end{eqnarray}
The equations for valley $K'$ can be obtained from the above by
replacing $U \to -U$. By imposing $\varepsilon=0$, we obtain the following bulk gap closing condition
\begin{equation}
M^2-U^2=0, \label{Eq:PhaseBoundaryforZeeman1}
\end{equation}
which has been reported in Ref.~[\onlinecite{James}]. In addition, if
$\lambda_R \neq \pm t_\perp$ and $U^2+t_\perp^2-\lambda_R^2\geq0$ then
a second gap closing condition
\begin{equation}
U^2+t_\perp^2-M^2-\lambda_R^2 = 0, \label{Eq:PhaseBoundaryforZeeman2}
\end{equation}
is also possible.
These two conditions signify two topological phase transitions and
give the corresponding phase transition boundaries.
\begin{figure}[!]
\includegraphics[width=7cm,angle=0]{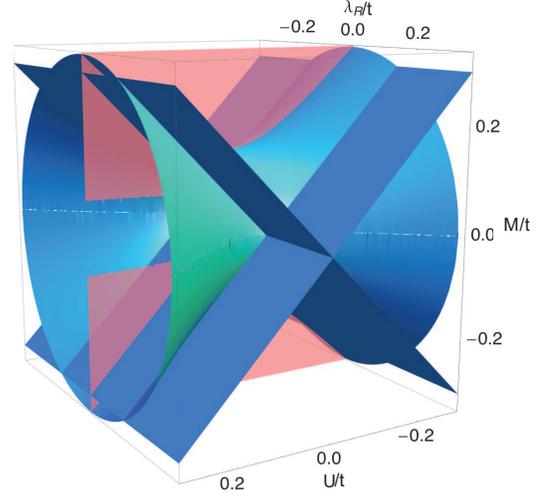}
\caption{(Color online) Phase diagram in the ($M$, $\lambda_R$, $U$) space. The whole phase space is divided into various topological phases. Here, we only show the phase boundaries. (1) The mutual vertical planes determined by Eq.~\eqref{Eq:PhaseBoundaryforZeeman1} separate the quantum valley Hall and quantum anomalous Hall phases. The uniparted hyperboloid determined by Eq.~{\eqref{Eq:PhaseBoundaryforZeeman2}} is used to separate different sub-phases. Note: in the plane of $\lambda_R=0$, the red color represents the metallic phase.}\label{Fig:FullPhase}
\end{figure}

\begin{figure*}
\includegraphics[width=18cm,angle=0]{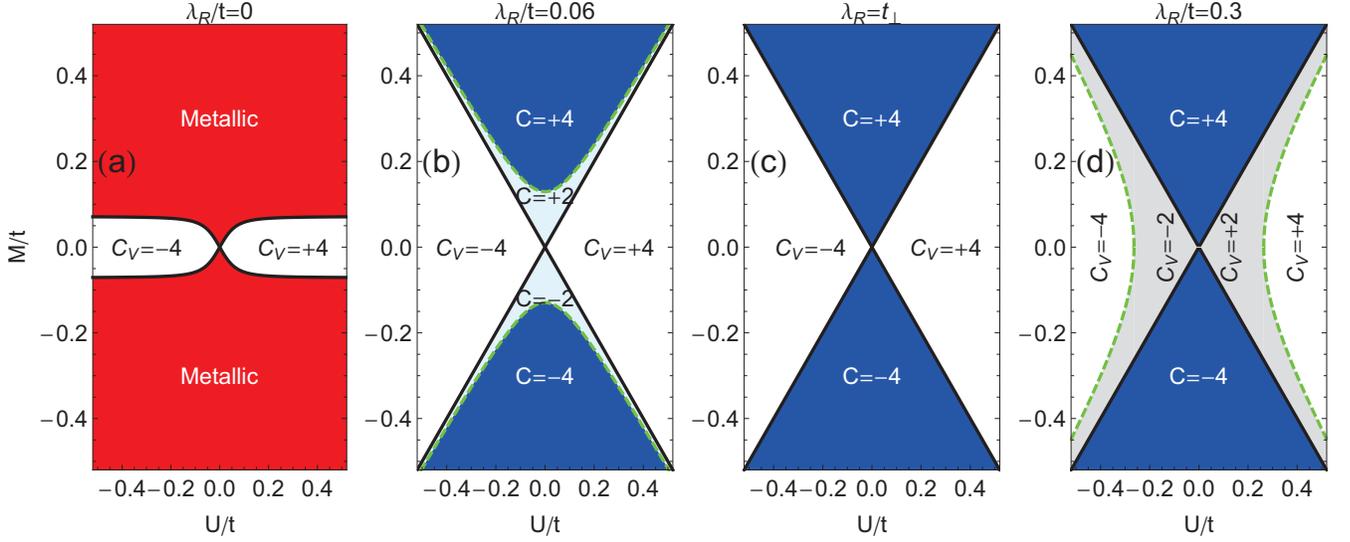}
\caption{(Color online) Phase diagram of bilayer graphene in the ($M$,
  $U$) plane at different fixed Rashba spin-orbit couplings
  $\lambda_R$. Solid lines represent the phase boundary
  $U=\pm M$, which separate the quantum valley-Hall region and the
  quantum anomalous-Hall region. The dashed lines are given by
  Eqs.~\eqref{Eq:PhaseBoundaryforZeeman1}-\eqref{Eq:PhaseBoundaryforZeeman2}, separating the two different
  quantum anomalous Hall phases [see panel (a)] or the two different quantum valley-Hall phases [see panel (c)]. (a) $\lambda_R<t_\perp$. the quantum valley-Hall phase is characterized by $\mathcal{C}_v=4 \sgn(U)$ (in white), while the quantum anomalous-Hall region comprises two different phases represented by $C=4 \sgn(M)$ (in blue/dark) and $C=2 \sgn(M)$ (in gray). (b) $\lambda_R=t_\perp$. There is no sub-phase in each region. The quantum valley-Hall phase and quantum anomalous-Hall phase are respectively characterized by $\mathcal{C}_v=4 \sgn(U)$ (in white) and $\mathcal{C}=4 \sgn(M)$ (in blue/dark). (c) $\lambda_R>t_\perp$. The quantum anomalous-Hall region has only one phase with Chern number being $\mathcal{C}=4 \sgn(M)$ (in blue/dark), while the quantum valley-Hall region includes two different phases with $\mathcal{C}_v=4 \sgn(U)$ (in white) and $\mathcal{C}_v=2\sgn(U)$ (in gray).}\label{Fig:2DQAHPhaseBoundary}
\end{figure*}

\begin{figure*}
\includegraphics[width=18cm,angle=0]{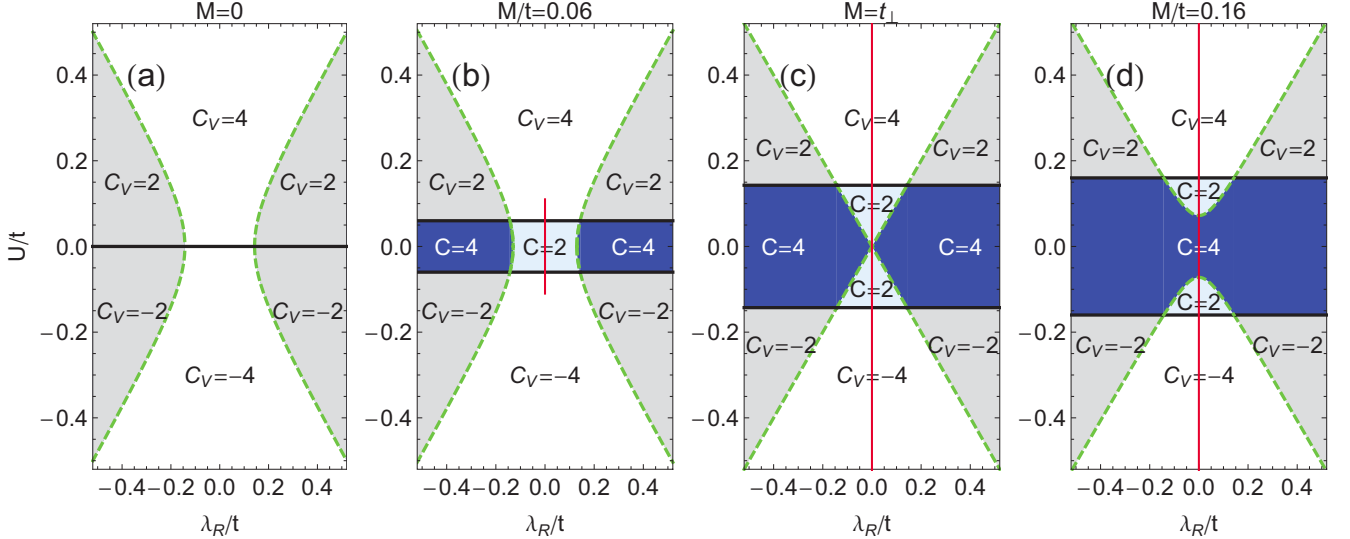}
\caption{(Color online) Phase diagram of bilayer graphene in the ($U$, $\lambda_R$) plane at different fixed exchange fields $M$. Solid lines represent the phase boundary $U=\pm M$, which separate the quantum valley-Hall region and the quantum anomalous-Hall region. The dashed lines represent phase boundary separating two different sub-phases in the same region, given by $U^2+t_\perp^2-M^2-\lambda_R^2 = 0$. (a) $M$=0. This is just a profile along $t_{\perp}/t = 0.1428$ in Fig.~\ref{wholephasediagram} to compare with panels (b) and (c). Red/dark regions are $Z_2$ topological insulator phase with $Z_2=1$ and $\mathcal{C}_v=2\sgn(U)$. The white region is the $\mathcal{C}_v=4 \sgn(U)$ quantum valley-Hall phase. (b) $M<t_\perp$. Inside the region of $|U|<|M|$, quantum anomalous-Hall phases emerge. For larger $\lambda_R$, the Chern number is $\mathcal{C}=4$, while for small $\lambda_R$ approaching zero, it becomes $\mathcal{C}=2$. Outside the region of $|U|<|M|$, the valley-Chern numbers remain the same as those without exchange field as shown in panel (a). (c) $M=t_\perp$. The phase boundary becomes linear $U^2-\lambda_R^2 =0$, which serves as a critical point changing the topology of the phase boundary. (d) $M>t_\perp$. The essential physics remains similar to that in panel (b).}\label{Fig:2DQAHPhaseBoundaryM}
\end{figure*}

In the above discussions, we have omitted a very important case in the
presence of vanishing Rashba SOC $\lambda_R=0$. It is known that a
large exchange field will close the bulk band gap induced from potential
difference and results in a metallic phase. Most importantly, the band
gap closing is not always exactly at the valley points as in other
phase transitions we have discussed. After rearranging the Hamiltonian, the $8\times8$ Hamiltonian can be written as:
\begin{eqnarray}
H_{K}= \left[
\begin{array}{cccc}
H_1 & T \\
T & H_2
\end{array}
\right],
\end{eqnarray}
where
\begin{eqnarray}
H_{1}= \left[
\begin{array}{cccc}
U+M & 0 & 0 & 0 \\
0 & U-M & 0 & 0 \\
0 & 0 & -U+M & 0 \\
0 & 0 & 0 & -U-M
\end{array}
\right],
\end{eqnarray}

\begin{eqnarray}
H_{2}= \left[
\begin{array}{cccc}
U-M & 0 & 0 & 0 \\
0 & U+M & 0 & 0 \\
0 & 0 & -U-M & 0 \\
0 & 0 & 0 & -U+M
\end{array}
\right],
\end{eqnarray}
and
\begin{eqnarray}
T= \left[
\begin{array}{cccc}
0 & v k_- & 0 & t_{\perp} \\
v k_+ & 0 & 0 & 0 \\
0 & 0 & 0 & v k_- \\
t_{\perp} & 0 & v k_+ & 0
\end{array}
\right].
\end{eqnarray}

Due to the particle-hole symmetry, the bulk gap closing must occur at $\varepsilon=0$. Therefore, the equation should satisfy the following
\begin{eqnarray}
&&U^2t^2_{\perp} + M^2 +(U^2-v^2 k^2)^2-M^2[t^2_{\perp}+2(U^2+v^2 k^2)] \nonumber \\
&&=0.
\end{eqnarray}
In order to have a real solution, we need
\begin{eqnarray}
(M^2+U^2)^2-(M^2-U^2)(M^2-U^2-t^2_{\perp})\geq 0,
\end{eqnarray}
which gives rise to the phase transition condition
\begin{eqnarray}
\frac{4}{t^2_{\perp}}=\frac{1}{M^2}-\frac{1}{U^2}.
\end{eqnarray}

In Fig.~\ref{Fig:FullPhase}, we provide a vivid three-dimensional (3D)
plot of the phase diagram in the ($M$, $\lambda_R$, $U$) space. For
clarity, we do not label each phase, but will distinguish them in the
subsequent 2D phase diagrams in detail. One can observe that the whole
3D space is divided by a set of mutual vertical planes and a uniparted
hyperboloid determined by Eqs.~\eqref{Eq:PhaseBoundaryforZeeman1} and
\eqref{Eq:PhaseBoundaryforZeeman2}, respectively. It is noteworthy
that the plane of $\lambda_R=0$ is distinct from other phase
boundaries, i.e., the region labeled as red is a metallic
phase. Below, we will explain the phase diagram by considering some
representative regions.

Figure~\ref{Fig:2DQAHPhaseBoundary} exhibits the phase diagrams in the
($U,M$) plane at four different fixed Rashba spin-orbit couplings: (a)
$\lambda_R=0$, (b) $\lambda_R<t_\perp$, (c) $\lambda_R=t_\perp$ and
(d) $\lambda_R>t_\perp$. In panel (a), one observes that for small $U$ the phase boundary is nearly linear to divide the metallic phase and quantum valley-Hall phase with valley Chern number $\mathcal{C}_v=4\sgn(U)$, while for larger $U$ the phase phase boundary becomes a constant. As can be seen from other three graphs in (b)-(d), the fundamental division of the parameter space into QVHI phase and quantum anomalous-Hall phase are separated by the solid lines given by $U=\pm M$. In our calculation, the total Chern number is defined by $\mathcal{C}=\mathcal{C}_{K}+\mathcal{C}_{K'}$. In Fig.~\ref{Fig:2DQAHPhaseBoundary}(b), the valley Chern number in the QVHI phase is $\mathcal{C}_v=4 \sgn(U)$ in the white regime, while the quantum anomalous-Hall region comprises two different phases of matter with Chern numbers being $\mathcal{C}=2 \sgn(M)$ and $\mathcal{C}=4 \sgn(M)$ denoted in gray and blue, respectively. When $\lambda_R=t_\perp$, the phase boundary is only determined by $U=\pm M$, which is the same as we discussed in Ref.~[\onlinecite{James}]. For a larger $\lambda_R$ as plotted in Fig.~\ref{Fig:2DQAHPhaseBoundary}(d), the Chern number in the quantum anomalous-Hall phase is $\mathcal{C}=4 \sgn(M)$, while the QVHI region includes two different phases characterized by valley Chern numbers $\mathcal{C}_v=2\sgn(U)$ and $\mathcal{C}_v=4 \sgn(U)$, represented in gray and white. It is interesting to point out that at fixed $M=0$ in the gray regime, it is both a $Z_2$ TI and a $\mathcal{C}_v=2\sgn(U)$ QVHI.

\begin{figure}[!]
\includegraphics[width=8cm,angle=0]{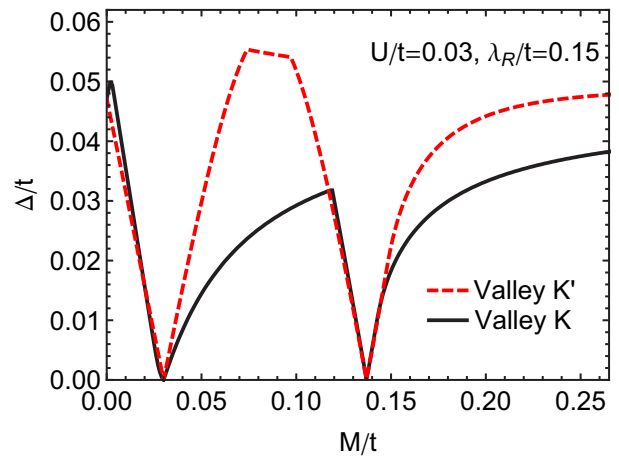}
\caption{(Color online) Bulk band gaps around valleys $K$ and $K'$ as
  a function of the exchange field $M$ at fixed $U=0.03t$ and
  $\lambda_R=0.15t$. Here, the interlayer hopping is set to be
  $t_\perp=0.1428t$. Solid and dashed lines represent the bulk band
  gap around $K$ and $K'$, respectively. It can be clearly seen that
  the two bulk gaps are unequal in general, except for the two
  critical points where the bulk gaps are completely
  closed. Consistent with the phase-diagram in
  Fig.~\ref{Fig:2DQAHPhaseBoundary}(c), it is a $\mathcal{C}_v=4$
  quantum valley-Hall insulator for small $M$ before the first bulk
  gap closing; when $M$ is located in the interval between two bulk
  gap closing points, the system enters into a $\mathcal{C}_v=2$
  quantum valley-Hall phase; for even larger $M$ exceeding the second critical point, it goes into a $\mathcal{C}=4$ quantum anomalous-Hall phase.}\label{Fig:DifferentK}
\end{figure}

In the above phase diagram, it is not obvious how the
$Z_2$ TI phase is affected by the presence of exchange
field. In Fig.~\ref{Fig:2DQAHPhaseBoundaryM}, we plot the phase
diagram in the ($U,\lambda_R$) plane at four fixed exchange fields:
(a) $M=0$; (b) $M<t_\perp$; (c) $M=t_\perp$; and (d)
$M>t_\perp$. Figure~\ref{Fig:2DQAHPhaseBoundaryM}(a) shows the phase
diagram in the absence of exchange field, which is the profile of
$t_\perp=0.1428t$ in Fig.~\ref{wholephasediagram}. We use gray and white colors to denote the $Z_2$
TI phase and conventional QVHI phase, respectively. When the exchange field is turned on, in
Figs.~\ref{Fig:2DQAHPhaseBoundaryM}(b)-(d), one finds that two
different quantum anomalous Hall phases with Chern numbers
$\mathcal{C}=2,4$ are induced when $U<M$ at nonzero Rashba effect. It is noteworthy that at
finite $M$ the ${Z}_2$ TI phase vanishes due to the
time-reversal symmetry breaking, but the valley Chern number remains
quantized $\mathcal{C}_v=2\sgn(U)$ when $|U|>M$. From these three
graphs, one observes that the phase boundary labeled by the dashed
lines are governed by Eq.~\eqref{Eq:PhaseBoundaryforZeeman2}, which reduces to the phase
boundary equation Eq.~\eqref{Eq:PhaseBoundary} in the limit of
$M=0$. This phase boundary indicates a continuity with and without
exchange field. One also observes that with increasing exchange field,
the topology of the phase boundary in dashed line changes at $M=t_\perp$. For zero Rashba SOC, when the exchange field is small, it closes the bulk band gap induced by small potential difference [see the vertical red line in panel (b)], driving the QVHI phase into a metallic phase; when the exchange field is large enough, the bulk gap from any potential difference is closed, giving rise to a complete metallic phase [see the red lines in panels (c) and (d)].

It is important to state that so far there are only a few papers~\cite{GuangyuGuo,JiangHua} that report tunable Chern numbers in a quantum anomalous Hall system. From the above analysis, it is clear that our system provides another platform that hosts quantum anomalous Hall phases with different Chern numbers.
The above phase diagrams are summarized concisely
in Table~\ref{Table:QuantumPhase}, which gives a complete
classification of all possible topological phases in the gated bilayer
graphene with Rashba spin-orbit coupling and exchange field, shown together
with the necessary conditions for a particular phase to occur.

\begin{table*}
  \caption{Summary of different topological phases in bilayer graphene in the presence of interlayer potential difference $U$, Rashba spin-orbit coupling $\lambda_R$, and exchange field $M$. They can be divided into two categories: quantum valley-Hall insulator (QVHI) and quantum anomalous-Hall insulator (QAHI). Note that the $Z_2$ TI also belongs to the QVHI phase.}
    \begin{tabular}{c|c|c|c|c|c}
   \hline\hline
          &      & $0< |\lambda_R| <t_{\perp}$ & $|\lambda_R|=t_\perp$ & \multicolumn{1}{c}{$|\lambda_R|>t_{\perp}$} \bigstrut[b]\\
    \hline
    \multirow{4}[8]{*}{QVHI} & \multirow{2}[4]{*}{$M=0$} & \multirow{2}[4]{*}{$\mathcal{C}_v=4 \sgn(U)$} & \multirow{2}[4]{*}{-}
    & \multicolumn{1}{l}{$U^2>\lambda_R^2-t_\perp^2$: $\mathcal{C}_v=4 \sgn(U)$} \bigstrut\\
\cline{5-5}          &       &       &       & \multicolumn{1}{l}{$U^2<\lambda_R^2-t_{\perp}^2$: $\mathcal{C}_v= 2\sgn(U)$ and $Z_2=1$} \bigstrut\\
\cline{2-5}          & \multirow{2}[4]{*}{$0<M^2<U^2$} & \multicolumn{2}{c|}{\multirow{2}[4]{*}{$\mathcal{C}_v=4\sgn(U)$}}
& \multicolumn{1}{l}{$M^2<U^2+t_{\perp}^2-\lambda_R^2$: $\mathcal{C}_v=4 \sgn(U)$} \bigstrut\\
\cline{5-5}          &       & \multicolumn{2}{c|}{} & \multicolumn{1}{l}{$M^2>U^2+t_{\perp}^2-\lambda_R^2$: $\mathcal{C}_v= 2\sgn(U)$} \bigstrut\\
    \hline
    \multirow{2}[3]{*}{QAHI} & \multirow{2}[3]{*}{$U^2<M^2$} & $M^2>U^2+t_\perp^2-\lambda_R^2$: $\mathcal{C}=4\sgn(M)$ & \multicolumn{2}{c}{\multirow{2}[3]{*}{$\mathcal{C}=4\sgn(M)$}} & \bigstrut\\
\cline{3-3}          &       & $M^2<U^2+t_\perp^2-\lambda_R^2$: $\mathcal{C}=2\sgn(M)$ & \multicolumn{2}{c}{} & \bigstrut[t]\\
\hline\hline
    \end{tabular}
  \label{Table:QuantumPhase}
\end{table*}

Another interesting feature in our system is that when all the three
parameters $U$, $M$ and $\lambda_R$ are nonzero, the bulk gaps at valleys
$K$ and $K'$ have different responses. As an example,
in Fig.~\ref{Fig:DifferentK} we present the results for the bulk band
gaps as a function of the exchange field $M$ at fixed interlayer
potential difference $U=0.03t$ and Rashba spin-orbit coupling
$\lambda_R=0.15t$. It is clearly seen that as long as the exchange
field term is turned on, the bulk gap amplitudes between K (solid
line) and K' (dashed line) become unequal. However, even though the
bulk gaps around the two valleys evolve quite differently, they close simultaneously. Again, the critical values of $M$ at the closing points agree very well with the analytic expression we derived in Eqs.~\eqref{Eq:PhaseBoundaryforZeeman1}-\eqref{Eq:PhaseBoundaryforZeeman2}.

\section{Summary}\label{section6}
We have derived low-energy Hamiltonian descriptions for the
TI phase and the QVHI phase in
$AB$-stacked bilayer graphene with interlayer potential $U$ and Rashba spin-orbit
coupling $\lambda_R$. We have explored the cases when the bilayer
graphene has an $AA$-stacking or has different Rashba spin-orbit
coupling strengths in the top and bottom layers. We showed that a
$Z_2$ TI state can only be realized in the
$AB$-stacked but not the $AA$-stacked bilayer graphene. To induce a
strong enough Rashba spin-orbit coupling in bilayer graphene, e.g., by
heavy metal dopants or a substrate,  different Rashba spin-orbit
coupling strengths in the top and bottom layers $\lambda_1 \neq
\lambda_2$ could arise. We find that the TI phase
can be realized as long as $\lambda_1 \lambda_2 > t^2_\perp$ for small
interlayer potential difference. When the time-reversal symmetry is
broken by an exchange field $M$, additional topological phases can be
induced. We find that the QVHI phase and quantum
anomalous-Hall phase are divided by $U=\pm M$. When $\lambda_R \neq
t_\perp$, there exists another topological phase boundary determined
by $U^2+t_\perp^2-M^2-\lambda_R^2=0$. For fixed $\lambda_R$, when
$\lambda_R < t_\perp$, the quantum anomalous-Hall phase contains two
different regions characterized by the Chern numbers of $C=2,4
\sgn(M)$; when $\lambda_R > t_\perp$, the QVHI phase separates into two regions characterized by the valley Chern numbers of $C_v=2,4 \sgn(U)$. Moreover, we find that when any two of the three parameters (i.e. interlayer potential difference, Rashba spin-orbit coupling, and the exchange field) are considered, the bulk band gaps at $K$ and $K'$ are equal. However, if all three terms are present, the bulk gaps at $K$ and $K'$ become different except at the topological phase transition points. It is noteworthy that in multilayer graphene, a bulk band gap opens in the presence of an external electric field. This makes multilayer graphene a good candidate to explore more interesting topological phases.~\cite{XiaoLi}

\emph{Acknowledgements.} This work was financially supported by
Welch Foundation (F-1255), DOE (DE-FG03-02ER45958, Division of
Materials Science and Engineering), the MOST Project of China
(2012CB921300), and NSFC (91121004). H.J. was supported by the CPSF
(20100480147 and 201104030). Y.Y. was supported by the NSF of China
(10974231 and 11174337) and the MOST Project of China (2011CBA00100).

\subsection{Appendix}

In the following discussion we set $\lambda_R$ and $t_\perp$ as fixed, and allow $U$ to slightly deviate from the phase transition point $U_0$, i.e., $U=U_0\pm \Delta$, where $U_0=\sqrt{\lambda^2_R - t^2_\perp}$ and $\pm$ correspond respectively to the quantum valley-Hall and topological insulator phases.

As stated in the main text, when the topological phase transition occurs, the bulk band gap closes at the exact Dirac points $K/K'$. At the critical $U_0$, the system Hamiltonian on the basis of $\{B_{1\downarrow}, A_{2\uparrow}, A_{1\uparrow}, A_{2\downarrow}, B_{2\uparrow}, B_{2\downarrow}, B_{1\uparrow}, A_{1\downarrow} \}$ can be expressed as
\begin{eqnarray}
H(U_0)=\left[
\begin{array}{cccccccc}
U_0 &  0   & 0       & 0           & 0          & 0       & 0           & 0 \\
0   & -U_0 & 0       & 0           & 0          & 0       & 0           & 0 \\
0   &  0   & U_0     & 0           & t_\perp    & 0       & 0           & 0 \\
0   &  0   & 0       & -U_0        & -i\lambda_R& 0       & 0           & 0 \\
0   &  0   & t_\perp & i \lambda_R & -U_0       & 0       & 0           & 0 \\
0   &  0   & 0       & 0           & 0          & -U_0    & 0           & t_\perp\\
0   &  0   & 0       & 0           & 0          & 0       &      U_0    & i\lambda_R\\
0   &  0   & 0       & 0           & 0          & t_\perp & -i\lambda_R & U_0
\end{array} \nonumber
\right].
\end{eqnarray}
Through a direct diagonalization, the eigenenergies are obtained as
$\varepsilon=0,0,\pm U_0, \pm \varepsilon_1, \pm \varepsilon_2$, where
$\varepsilon_1=- U_0/2 + \sqrt{8\lambda^2_R + U^2_0}/2$ and
$\varepsilon_2=- U_0/2 - \sqrt{8\lambda^2_R + U^2_0}/2$. The former
four correspond to the low-energy part, while the latter four
correspond to the high-energy part. Based on our analysis of the Berry
curvature from the tight-binding model, the valley-Chern number arises
only from the low-energy bands and the high-energy bands contribute
zero.

For the diagonal block Hamiltonian of
$h_1(U_0)=\left[
\begin{array}{cccccccc}
 U_0     & 0           & t_\perp     \\
 0       & -U_0        & -i\lambda_R \\
 t_\perp & i \lambda_R & -U_0        \\
\end{array} \nonumber
\right]
$, its unitary transformation matrix is
\begin{eqnarray}
\mathbf{V}_1&=&\left[
\begin{array}{cccccccc}
 v_{11} & v_{12} & v_{13} \\
 v_{21} & v_{22} & v_{23} \\
 v_{31} & v_{32} & v_{33} \\
\end{array} \nonumber
\right] \\
&=&\left[
\begin{array}{cccccccc}
 \frac{t_\perp}{\sqrt{2} \lambda_R} & {1}/{\alpha_2} & {1}/{\alpha_3} \\
 \frac{i}{\sqrt{2}} & \frac{i \lambda_R (U_0-\varepsilon_1)}{\alpha_2 t_\perp (U_0 +\varepsilon_1)} & \frac{i \lambda_R (U_0-\varepsilon_2)}{\alpha_3 t_\perp (U_0 +\varepsilon_2)} \\
 \frac{-U_0}{\sqrt{2} \lambda_R} & \frac{\varepsilon_1-U_0}{\alpha_2 t_\perp} & \frac{\varepsilon_2-U_0}{\alpha_3 t_\perp} \\
\end{array} \nonumber
\right],
\end{eqnarray}
which can diagonalize $h_1(U_0)$ to be
\begin{eqnarray}
h^\prime_1(U_0)=\mathbf{V}^\dag_1 h_1(U_0)\mathbf{V}_1=\left[
\begin{array}{cccccccc}
 0 & 0 & 0 \\
 0 & \varepsilon_1 & 0 \\
 0 & 0 & \varepsilon_2 \\
\end{array} \nonumber
\right].
\end{eqnarray}
$\alpha_2$ and $\alpha_3$ are respectively
\begin{eqnarray}
\alpha_2&=&\sqrt{1+[\frac{\lambda_{\rm R}(U_0-\varepsilon_1)}{(U_0+\varepsilon_1)t_{\bot}}]^2+[\frac{\varepsilon_1-U_0}{t_{\bot}}]^2}, \\ \alpha_3&=&\sqrt{1+[\frac{\lambda_{\rm R}(U_0-\varepsilon_2)}{(U_0+\varepsilon_2)t_{\bot}}]^2+[\frac{\varepsilon_2-U_0}{t_{\bot}}]^2}.
\end{eqnarray}

For the other diagonal block Hamiltonian of
$h_2(U_0)=\left[
\begin{array}{cccccccc}
 -U_0     & 0           & t_\perp     \\
 0       & U_0        & i\lambda_R \\
 t_\perp & -i \lambda_R & U_0        \\
\end{array} \nonumber
\right],
$ its unitary transformation matrix is written as
\begin{eqnarray}
\mathbf{V}_2=\left[
\begin{array}{cccccccc}
 v_{11} & v_{12} & v_{13} \\
 -v_{21} & -v_{22} & -v_{23} \\
 -v_{31} & -v_{32} & -v_{33} \\
\end{array} \nonumber
\right], \\
\end{eqnarray}
which leads to
\begin{eqnarray}
h^\prime_2(U_0)=\mathbf{V}^\dag_2 h_2(U_0)\mathbf{V}_2=\left[
\begin{array}{cccccccc}
 0 & 0 & 0 \\
 0 & -\varepsilon_1 & 0 \\
 0 & 0 & -\varepsilon_2 \\
\end{array} \nonumber
\right].
\end{eqnarray}

In order to arrange the eigenenergies to be low and high energy parts, the basis should be reordered to be: $\{A_{1\uparrow}, B_{2\downarrow}, B_{1\downarrow}, A_{2\uparrow}, A_{2\downarrow}, B_{2\uparrow}, B_{1\uparrow}, A_{1\downarrow}\}$. The corresponding full unitary transformation matrix becomes:
\begin{eqnarray}
\mathbf{V}=\left[
\begin{array}{cccccccc}
0 & 0 & 1 & 0 & 0 & 0 & 0 & 0 \\
0 & 0 & 0 & 1 & 0 & 0 & 0 & 0 \\
v_{11} & 0 & 0 & 0 & v_{12} & v_{13} & 0 & 0 \\
v_{21} & 0 & 0 & 0 & v_{22} & v_{23} & 0 & 0 \\
v_{31} & 0 & 0 & 0 & v_{32} & v_{33} & 0 & 0 \\
0 & v_{11} & 0 & 0 & 0 & 0 & v_{12} & v_{13} \\
0 & -v_{21} & 0 & 0 & 0 & 0 & -v_{22} & -v_{23} \\
0 & -v_{31} & 0 & 0 & 0 & 0 & -v_{32} & -v_{33} \\
\end{array} \nonumber
\right].
\end{eqnarray}

When the interlayer potential difference is slightly deviated from $U_0$, the Hamiltonian is written as:
\begin{eqnarray}
H(U)=H(U_0)+H_{\Delta}+H_{\bm k},
\end{eqnarray}
where
\begin{eqnarray}
H_{\Delta}=\text{Diag}\{\Delta, -\Delta, \Delta, -\Delta, -\Delta,-\Delta, \Delta, \Delta\}.
\end{eqnarray}
and
\begin{eqnarray}
H_{\bm k}=\left[
\begin{array}{cccccccc}
0 & 0 & 0 & 0 & 0 & 0 & 0 & v k_+ \\
0 & 0 & 0 & 0 & v k_- & 0 & 0 & 0 \\
0 & 0 & 0 & 0 & 0 & 0 & v k_- & 0 \\
0 & 0 & 0 & 0 & 0 & v k_- & 0 & 0 \\
0 & v k_+ & 0 & 0 & 0 & 0 & 0 & 0 \\
0 & 0 & 0 & v k_+ & 0 & 0 & 0 & 0 \\
0 & 0 & v k_+ & 0 & 0 & 0 & 0 & 0 \\
v k_- & 0 & 0 & 0 & 0 & 0 & 0 & 0 \\
\end{array} \nonumber
\right].
\end{eqnarray}

By performing a unitary transformation, the Hamiltonian of $H(U)$ becomes
\begin{widetext}
\begin{eqnarray}
&H'(U)&=\mathbf{V}^\dagger H(U) \mathbf{V} \nonumber \\
&=&\left[
\begin{array}{cccccccc}
(2v^2_{11}-1)\Delta & \gamma_1 v k_- & 0 & v_{31}v k_+ & 2v_{11}v_{12}\Delta & 2v_{11}v_{13}\Delta & \gamma_2 v k_- & \gamma_3 v k_- \\
-\gamma_1 v k_+ & (1-2v^2_{11})\Delta & -v_{31}v k_- & 0 & -\gamma_2 v k_+ & -\gamma_3 v k_+ & -2v_{11}v_{12}\Delta & -2v_{11}v_{13}\Delta \\
0 & -v_{31}v k_+ & \Delta+U_0 & 0 & 0 & 0 & -v_{32} v k_+ & -v_{33} v k_+ \\
v_{31}v k_- & 0 & 0 & -\Delta-U_0 & v_{32} v k_- & v_{33} v k_- & 0 & 0 \\
2v_{11}v_{12}\Delta & \gamma_2 v k_- & 0 & v_{32} v k_+ & \gamma_4*\Delta+ \varepsilon_1 & \gamma_5 \Delta & -2v_{12}v_{22}v k_- & \gamma_8 v k_- \\
2v_{11}v_{13}\Delta & \gamma_3 v k_- & 0 & v_{33} v k_+ & \gamma_5 \Delta & \gamma_6 \Delta+\varepsilon_2 & \gamma_8 v k_- & -2v_{13}v_{23}v k_- \\
-\gamma_2 v k_+ & -2v_{11}v_{12}\Delta & -v_{32} v k_- & 0 & 2v_{12}v_{22}v k_+ & -\gamma_8 v k_+ & -\gamma_4*\Delta-\varepsilon_1 & \gamma_7 \Delta \\
-\gamma_3 v k_+ & -2v_{11}v_{13}\Delta & -v_{33} v k_- & 0 & -\gamma_8 v k_+ & 2v_{13}v_{23}v k_+ & \gamma_7 \Delta & -\gamma_6 \Delta-\varepsilon_2 \\
\end{array} \nonumber
\right] \\
&=&\left[
\begin{array}{cccccccc}
H_P& T \\
T^\dagger & H_Q \\
\end{array} \nonumber
\right], \\
\end{eqnarray}
\end{widetext}
where $\gamma_1=-2 v_{11} v_{21}=-i t_\perp / \lambda_R$, $\gamma_2=-(v_{11} v_{22} + v_{12} v_{21})$, $\gamma_3=-(v_{11}v_{23}+v_{13}v_{21})$, $\gamma_4=v^2_{12}+v^2_{22}-v^2_{32}$, $\gamma_5=v_{12}v_{13}+v_{22}v_{23}-v_{32}v_{33}$, $\gamma_6=v^2_{13}+v^2_{23}-v^2_{33}$, $\gamma_7=v_{32}v_{33}-v_{22}v_{23}-v_{12}v_{13}$, and $\gamma_8=-(v_{12}v_{23}+v_{13}v_{22})$. Since both $\Delta$ and $k_+/h_-$ are extremely small, the effective Hamiltonian can be simplified to be
\begin{eqnarray}
H_{eff}=H_P-T H^{-1}_Q T^{\dagger}.
\end{eqnarray}

Explicitly, $H_P$ can be written as:
\begin{eqnarray}
H_P=\left[
\begin{array}{cccccccc}
\frac{-U^2_0 \Delta}{\lambda^2_R}   & \frac{-it_\perp v k_-}{\lambda_R} & 0 & \frac{-U_0 v k_+}{\sqrt{2}\lambda_R} \\
\frac{it_\perp v k_+}{\lambda_R} & \frac{U^2_0 \Delta}{\lambda^2_R}  & \frac{U_0 v k_-}{\sqrt{2}\lambda_R} & 0\\
0 & \frac{U_0 v k_+}{\sqrt{2}\lambda_R} & \Delta+U_0 & 0  \\
\frac{-U_0 v k_-}{\sqrt{2}\lambda_R} & 0 & 0 & -\Delta-U_0 \\
\end{array} \nonumber
\right]. \\
\end{eqnarray}

For small $\Delta$ and ${\bm k}$, the higher energy block can be expressed as:
\begin{eqnarray}
H_Q=\left[
\begin{array}{cccccccc}
\varepsilon_1 & 0 & 0 & 0 \\
0&\varepsilon_2&0&0 \\
0&0&-\varepsilon_1&0 \\
0&0&0&-\varepsilon_2 \\
\end{array} \nonumber
\right]. \\
\end{eqnarray}


\begin{thebibliography}{99}
\bibitem{latticevector}
The lattice unit vectors are the same as those in Ref.~[\onlinecite{qiao3}].
\bibitem{note1}
For details of the expression of each term, one can refer to Ref.~[\onlinecite{qiao3}].
\bibitem{note0}
This is reasonable, because based on our numerical results the high-energy valence bands have respectively opposite Berry curvatures making no contribution to the total Berry curvatures.
\bibitem{note2}
The corresponding band structures from the model Hamiltonian at $K'$ are exactly the same as those from valley $K$.
\end{thebibliography}
\end{document}